\documentclass[conference,screen]{IEEEtran}
\IEEEoverridecommandlockouts
\usepackage{cite}
\usepackage{amsmath,amssymb,amsfonts}
\usepackage{graphicx}
\usepackage{textcomp}
\usepackage{xcolor}
\usepackage{booktabs}
\usepackage[protrusion=true,expansion=true]{microtype}
\usepackage{listings}
\definecolor{codekw}{rgb}{0.00,0.20,0.70}
\definecolor{codecm}{rgb}{0.00,0.50,0.20}
\definecolor{codest}{rgb}{0.58,0.00,0.55}
\definecolor{codebg}{rgb}{0.97,0.97,0.97}
\definecolor{coderl}{rgb}{0.55,0.55,0.55}
\lstdefinestyle{tsstyle}{
  language=Java,
  basicstyle=\ttfamily\footnotesize,
  keywordstyle=\color{codekw}\bfseries,
  commentstyle=\color{codecm}\itshape,
  stringstyle=\color{codest},
  backgroundcolor=\color{codebg},
  rulecolor=\color{coderl},
  frame=lines,
  framerule=0.3pt,
  framesep=2pt,
  columns=fullflexible,
  keepspaces=true,
  breaklines=false,
  showstringspaces=false,
  morekeywords={await,async,let,from,of},
  xleftmargin=0.5em,
  aboveskip=0.5em,
  belowskip=0.3em,
  captionpos=b,
  abovecaptionskip=0.4em,
  belowcaptionskip=0.4em,
}
\usepackage[hidelinks]{hyperref}
\usepackage{orcidlink}

\def\BibTeX{{\rm B\kern-.05em{\sc i\kern-.025em b}\kern-.08em
    T\kern-.1667em\lower.7ex\hbox{E}\kern-.125emX}}

\begin{document}

\title{When EVM Defaults Break: \\ Three SDK Patterns for UTXO-Family Wallets}

\author{
\IEEEauthorblockN{
Hanzheng Zhang\,\orcidlink{0009-0001-9726-6168}
}
\IEEEauthorblockA{
Common Knowledge Base Association\\
China\\
\href{mailto:0@hanssen0.com}{0@hanssen0.com}
}
\and
\IEEEauthorblockN{
Hongzhou Chen\,\orcidlink{0000-0002-8310-3202}\textsuperscript{*}
}
\IEEEauthorblockA{
Common Knowledge Base Association\\
Baoliang Education (HK) Group Ltd.\\
Hong Kong SAR\\
\href{mailto:hongzhouchen@ieee.org}{hongzhouchen@ieee.org}\\
}
\thanks{\textsuperscript{*}Corresponding author: Hongzhou Chen.}
}

\maketitle

\begin{abstract}
Mainstream Web3 SDKs (e.g., ethers.js, viem) bundle three coupled assumptions: an account-based ledger, a single secp256k1 signer, and input-selection-free transaction construction. These assumptions fail once a chain adopts an extended-UTXO model (here, the Cell model on Nervos CKB), and once the wallet population spans heterogeneous cryptographic primitives (secp256k1, WebAuthn, Schnorr). Using CKBers' Codebase (CCC), a TypeScript SDK, this retrospective case study extracts three recurring patterns from release history and exported APIs: a \emph{Transaction Composer} that makes capacity-aware input completion and fee estimation declarative, a \emph{Unified Signer Abstraction} that normalizes signing across heterogeneous wallets behind one interface, and a \emph{Connector UI} component that ships per-wallet onboarding as a drop-in. The contribution is the articulation of these patterns as transferable design lessons for SDKs targeting UTXO-family ledgers and heterogeneous wallet populations.
\end{abstract}

\begin{IEEEkeywords}
Web3 SDK, wallet abstraction, extended UTXO, key management, design patterns.
\end{IEEEkeywords}

\section{Introduction}

The dominant client-side Web3 toolchain is shaped by Ethereum. Libraries such as ethers.js and viem encode three load-bearing defaults~\cite{wood2014ethereum,ethersjs,viem}: state lives in accounts rather than unspent outputs~\cite{nakamoto2008bitcoin}, a single secp256k1 key behind an externally owned account (EOA) signs every transaction, and a transaction is a nonce-ordered payload constructed without selecting concrete inputs. On extended-UTXO chains, these defaults fail together. \textbf{Extended-UTXO chains}~\cite{chakravarty2020eutxo}, including the Cell model of Nervos CKB~\cite{nervos2019whitepaper} studied here, place state inside typed UTXO-like units (cells) guarded by user-defined scripts, with capacity rather than gas as the scarce resource. Such chains also serve a heterogeneous wallet population: deployed CKB wallets sign with secp256k1 ECDSA, WebAuthn passkeys, or Schnorr keys (Nostr), each with its own signing path and key-management model~\cite{eskandari2018keymgmt,krombholz2016ux}. Through CKBers' Codebase (\href{https://github.com/ckb-devrel/ccc}{CCC}), a TypeScript SDK with over 330 tagged releases, this poster identifies and contributes three reusable design patterns.

\section{Background: UTXO-Family Wallet Constraints}

Three features of the extended-UTXO setting make wallet-side SDK design qualitatively different from the EVM case. First, every UTXO-like unit (a cell) must carry enough \emph{capacity} to cover the byte size of its serialized output and data, so transaction construction requires balancing inputs and outputs against capacity, not only value, and producing a change unit. Second, the chain is script-pluralistic: a lock script can encode any signature scheme, so different unit owners require different signing pipelines even within one transaction~\cite{nervos2019whitepaper}. Third, fee estimation is coupled to transaction byte size and lock-script choice, a feedback loop absent from the EVM gas model.

\begin{lstlisting}[float=b,caption={A demo transfer with CCC. Three awaited calls complete inputs by capacity, set the fee, and sign and send.},label={lst:transfer},abovecaptionskip=0.4em]
// Transfer 100 CKB to `receiver' with the SDK
const tx = ccc.Transaction.from({
  outputs: [{lock: receiver,
            capacity: ccc.fixedPointFrom(100)}],});
await tx.completeInputsByCapacity(signer);
await tx.completeFeeBy(signer);
await signer.sendTransaction(tx);
\end{lstlisting}

\begin{table*}[!t]
\caption{Three SDK patterns for UTXO-family wallet interaction, as instantiated in CCC.}
\label{tab:patterns}
\centering
\footnotesize
\renewcommand{\arraystretch}{0.9}
\begin{tabular}{@{}p{2cm}p{3.3cm}p{3.8cm}p{3.9cm}p{3.6cm}@{}}
\toprule
\textbf{Pattern} & \textbf{EVM-centric assumption} & \textbf{UTXO-family requirement} & \textbf{CCC mechanism} & \textbf{User-facing effect} \\
\midrule
A. TX Composer & Caller supplies to/value/data, SDK fills nonce and gas. & Inputs, change units, byte size, and fee are mutually coupled. & Declarative outputs, SDK resolves inputs, change, and fee in one pass. & No manual capacity iteration in application code. \\
\midrule
B. Unified Signer & Single secp256k1 signer behind an EOA. & secp256k1, WebAuthn, and Schnorr coexist on one chain. & One \texttt{Signer} interface encapsulating per-scheme differences. & One codepath drives eight heterogeneous wallet adapters. \\
\midrule
C. Connector UI & DApps re-implement wallet detection and onboarding. & Per-wallet detection, chain switching, and account-change handling. & Drop-in Web Component or React wrapper around the Signer. & Onboarding cost paid once by the SDK, not per DApp. \\
\bottomrule
\end{tabular}
\end{table*}

\section{Three Design Patterns in CCC}

\begin{figure}[t]
\centering
\includegraphics[width=\columnwidth]{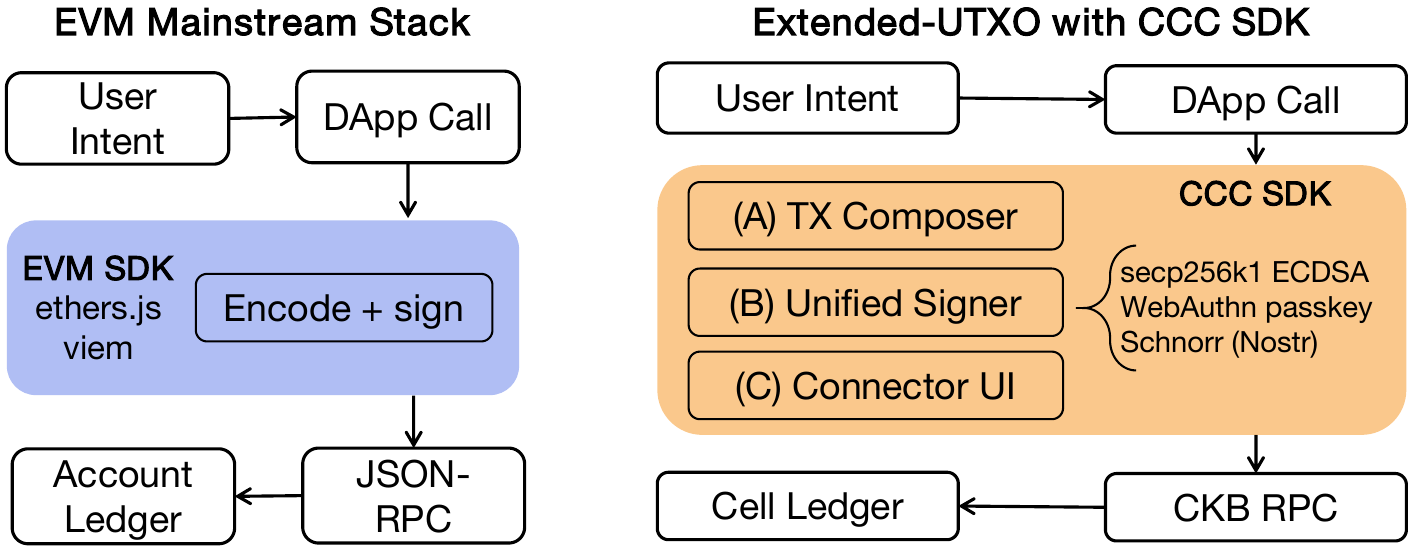}
\caption{EVM SDK versus CCC SDK on extended-UTXO chains. Patterns A, B, and C are co-resident modules.}
\label{fig:contrast}
\end{figure}

Fig.~\ref{fig:contrast} contrasts the EVM default stack with the CCC stack, Table~\ref{tab:patterns} summarizes the three patterns, and Listing~\ref{lst:transfer} shows the resulting application-side code.

\subsection{Pattern A: Transaction Composer}

In an EVM SDK, transaction construction centers on the user's intent, without selecting concrete spendable inputs. By contrast, to enable concurrent transaction execution, cell-model transactions require first resolving input units, change, and the byte-size-dependent fee off-chain. CCC absorbs this coupling behind a Composer that accepts declarative outputs and resolves inputs, change, and fee in one pass, removing manual capacity arithmetic from application code. The cost of converging on this abstraction is visible in the history of \texttt{@ckb-ccc/core}: between July 2024 and August 2025 the fee-handling surface grew into seven interrelated methods, including variants that route change to a new lock or an existing output.

\subsection{Pattern B: Unified Signer Abstraction}

Each signing scheme on the chain carries a different message-construction convention, witness encoding, and key-management model~\cite{eskandari2018keymgmt,krombholz2016ux}. CCC exposes a single \texttt{Signer} interface whose concrete implementations encapsulate differences, so downstream protocols see every wallet as the same object. Across eight wallet adapters, this normalizes three signing primitives: secp256k1 ECDSA (covering Bitcoin, CKB, EVM, and Dogecoin conventions), WebAuthn passkeys, and Schnorr keys (Nostr). The JoyID adapter alone drives CKB, Bitcoin, EVM, and Nostr signing behind one interface. This is capability normalization, placing scheme-specific method responsibility inside adapter implementations rather than in application code~\cite{stylos2008apiusability}. The rejected alternative, exposing per-scheme signer classes as Bitcoin toolchains do, would push scheme dispatch into every application. The accepted cost is a lowest-common-denominator interface, with scheme-specific capabilities surfaced through optional extensions.

\subsection{Pattern C: Connector UI as Drop-in Infrastructure}

Heterogeneous-wallet onboarding requires detecting the provider, requesting permissions, handling account changes, and rendering a dialog. CCC ships this surface as a reusable Web Component (\texttt{@ckb-ccc/connector}) and a React wrapper (\texttt{@ckb-ccc/connector-react}), so the DApp declares which wallets to enable. This treats wallet connection as \emph{infrastructure} whose per-wallet cost is paid once by the SDK. The alternative of leaving onboarding to third-party kits, as EVM DApps do with RainbowKit-style libraries, was unavailable for this wallet population, and the accepted cost is that the SDK itself must track per-wallet API churn. Adoption evidence for this pattern is outside this paper's data sources.

\section{Discussion and Transferability}

This is a retrospective case study grounded in the SDK's public release history and exported API surface. As of the data-collection cutoff of June 10, 2026, the artifact comprises 18 published \texttt{@ckb-ccc/*} packages on npm, including the eight wallet adapters above, 13 human contributors, and roughly 190 independent \texttt{package.json} files on GitHub listing \texttt{@ckb-ccc/core} as a dependency.\footnote{GitHub code-search count, June 10, 2026. Approximate, and excludes the repository's own workspace files.} Patterns were identified by tracing which exported API shapes stabilized and recurred from July 2024 to June 2026, keeping those spanning multiple wallets or transaction types rather than one-off helpers. These patterns generalize to other extended-UTXO ecosystems~\cite{chakravarty2020eutxo}, most prominently Cardano, where declarative composition and reusable connectors already appear. The more distinctive contribution is therefore Pattern B's unification across signing primitives, consistent with calls to lift wallet abstraction out of the application layer. The same heterogeneity is arriving on EVM through account abstraction (ERC-4337), so the lesson applies beyond UTXO-family chains. Analytic transfer holds when transaction validity requires explicit UTXO-family resource balancing and wallet access spans more than one signing primitive. The central lesson is that an SDK in this regime should hide routine complexity by default while keeping the manual path open. In CCC, the completion calls are optional helpers over a transaction object that applications can still assemble field by field. Future work should test whether these patterns reduce integration time or defects.

\bibliographystyle{IEEEtran}
\bibliography{ref}

@misc{nakamoto2008bitcoin,
  author       = {Nakamoto, Satoshi},
  title        = {Bitcoin: A Peer-to-Peer Electronic Cash System},
  year         = {2008},
  howpublished = {\url{https://bitcoin.org/bitcoin.pdf}}
}

@techreport{wood2014ethereum,
  author      = {Wood, Gavin},
  title       = {Ethereum: A Secure Decentralised Generalised Transaction Ledger},
  institution = {Ethereum Project},
  type        = {Yellow Paper},
  year        = {2014}
}

@misc{nervos2019whitepaper,
  author       = {Xie, Jan},
  title        = {Nervos {CKB}: A Common Knowledge Base for {Crypto-Economy}},
  year         = {2018},
  howpublished = {\url{https://github.com/nervosnetwork/rfcs/blob/master/rfcs/0002-ckb/0002-ckb.md}},
  note         = {{RFC-0002} white paper}
}

@inproceedings{krombholz2016ux,
  author    = {Krombholz, Katharina and Judmayer, Aljosha and Gusenbauer, Matthias and Weippl, Edgar},
  title     = {The Other Side of the Coin: User Experiences with {Bitcoin} Security and Privacy},
  booktitle = {Financial Cryptography and Data Security (FC)},
  series    = {LNCS},
  volume    = {9603},
  pages     = {555--580},
  year      = {2016},
  publisher = {Springer},
  doi       = {10.1007/978-3-662-54970-4_33}
}

@inproceedings{eskandari2018keymgmt,
  author    = {Eskandari, Shayan and Barrera, David and Stobert, Elizabeth and Clark, Jeremy},
  title     = {A First Look at the Usability of {Bitcoin} Key Management},
  booktitle = {Proc.\ NDSS Workshop on Usable Security (USEC)},
  year      = {2015},
  publisher = {Internet Society},
  doi       = {10.14722/usec.2015.23015}
}

@inproceedings{chakravarty2020eutxo,
  author    = {Chakravarty, Manuel M. T. and Chapman, James and MacKenzie, Kenneth and Melkonian, Orestis and Peyton Jones, Michael and Wadler, Philip},
  title     = {The Extended {UTXO} Model},
  booktitle = {Financial Cryptography and Data Security (FC) Workshops},
  series    = {LNCS},
  volume    = {12063},
  pages     = {525--539},
  year      = {2020},
  publisher = {Springer},
  doi       = {10.1007/978-3-030-54455-3_37}
}

@inproceedings{stylos2008apiusability,
  author    = {Stylos, Jeffrey and Myers, Brad A.},
  title     = {The Implications of Method Placement on {API} Learnability},
  booktitle = {Proc.\ ACM SIGSOFT Symp.\ Foundations of Software Engineering (FSE)},
  pages     = {105--112},
  year      = {2008},
  publisher = {ACM},
  doi       = {10.1145/1453101.1453117}
}

@misc{ethersjs,
  author       = {Moore, Richard},
  title        = {ethers.js: Complete {Ethereum} Library and Wallet Implementation in {JavaScript}},
  year         = {2024},
  howpublished = {\url{https://docs.ethers.org/}}
}

@misc{viem,
  author       = {Moxey, Jake and Meagher, Tom},
  title        = {viem: {TypeScript} Interface for {Ethereum}},
  year         = {2024},
  howpublished = {\url{https://viem.sh/}}
}

\end{document}